\documentstyle[preprint,aps]{revtex}
\begin{document}   
\baselineskip 24pt     

\begin{center}   
{\Large \bf A New Model-independent Method of Determining 
$|V_{ub}/V_{cb}|$} 

\vspace{1.5cm}
 
{\Large C. S. Kim}\\

{\it Department of Physics, Yonsei University,
Seoul 120-749, Korea}\footnote{kim@cskim.yonsei.ac.kr, cskim@kekvax.kek.jp,
                              (fax) +82-2-392-1592}\\

\vspace{0.8cm}

{\bf Abstract} 
\end{center}

In order to determine the ratio of CKM matrix elements 
$|V_{ub}/V_{cb}|$ ~(and $|V_{ub}|$), we propose a new model-independent 
method based on the heavy quark effective theory, 
which is theoretically described by the phase space factor
and the well-known perturbative QCD correction only.
In the forthcoming asymmetric $B$-experiments with microvertex 
detectors, BABAR and BELLE, the total separation of $b \rightarrow u$ 
semileptonic decays from the dominant 
$b \rightarrow c$ semileptonic decays would be experimentally viable.
We explore the possible experimental option: the measurement of
inclusive hadronic invariant mass distributions. 
We also clarify the relevant experimental backgrounds.

\vspace{0.5cm}


\vfill

\pagebreak 

\baselineskip 22pt

\noindent
{\bf 1. Introduction}~~~
A precise determination of Cabibbo-Kobayashi-Maskawa (CKM) matrix
elements \cite{ckm} is the most important goal of the forthcoming
asymmetric $B$-factories \cite{B-fact}, KEKB and SLACB.
Their precise values are urgently needed for analyzing 
CP-violation and for
testing the Standard Model (SM) through the unitarity relations
among them. Furthermore, the accurate knowledge 
of these matrix elements can
be useful in relating them to the fermion masses and also 
in the searches for hints of new physics beyond the SM.

The CKM matrix element $V_{ub}$ is important to the SM description  
of CP-violation. If it were zero, there would be no CP-violation from 
the CKM matrix ({\it i.e.} in the SM), and we have to seek 
for other sources of CP violation in $K_{L} \rightarrow \pi\pi$. 
Observations of semileptonic $b\rightarrow u$ transitions by the CLEO 
\cite{cleo} and ARGUS \cite{argus}  
imply that $V_{ub}$ is  indeed nonzero, 
and it is important to extract the modulus $|V_{ub}|$ from 
semileptonic decays of $B$ mesons as accurately as possible.  
  
Historically, the charged lepton energy spectrum ($d \Gamma / d E_l$)
has been measured, and the $b\rightarrow u$ 
events are selected from the high 
end of the charged lepton energy spectrum.  
However, this cut on $E_l$ is not very effective, 
since only less than
$10 \%$ of $b\rightarrow u$ events\footnote{If it were not 
for the theoretical uncertainty, this $E_l$ cut would be very 
effective -- it completely suppresses
the $b \rightarrow c$ background.} 
survive this cut at the $B$ meson rest frame. 
(In the future asymmetric $B$-factories with boosted $B$ mesons, 
much less than $10 \%$ of $b\rightarrow u$ events would 
survive the $E_l$ cut over the $b \rightarrow c$ threshold.)
We also note that the dependences of  the lepton energy spectrum  on 
perturbative and non-perturbative QCD corrections \cite{kuhn,hqet} 
as well as on the unavoidable specific model parameters 
({\it e.g.} the parameter $p_{_F}$ of the ACCMM model \cite{accmm}) 
are strongest at the end-point region, 
which makes the model-independent determination of $|V_{ub}/V_{cb}|$ 
almost impossible from the inclusive distribution of
$d \Gamma / d E_l$.
For exclusive $B \rightarrow X_u l \nu$ decays, the application of 
heavy quark effective theory (HQET) is very much limited, 
since $u$-quark is not heavy compared to $\Lambda_{QCD}$. 
And the theoretical predictions
for the required hadronic matrix elements are quite
different depending on which model we use \cite{models}. 
However, in the long run theoretical uncertainties on $V_{ub}$ from 
the exclusive form factors are possibly reduced 
to the $10 \sim 15$\% level through the measurements on $q^2$ 
dependence of the form factors,  
and the exclusive semileptonic decays will also be providing 
valuable information.
By using the neutrino reconstruction technique 
and the beam constrained
invariant mass, CLEO \cite{alexander} has recently succeeded
measuring the branching ratio, 
${\cal{B}}(B^0 \rightarrow \rho^- l^+ \nu)=
(2.5 \pm 0.4^{+0.5}_{-0.7}\pm 0.5)
\times 10^{-4}$, where the errors are statistical, 
systematic and the estimated model-dependence based on the spread 
of models and individual
model errors. And they estimated 
$|V_{ub}|=(3.3 \pm 0.2^{+0.3}_{-0.4} \pm 0.7) \times 10^{-3}$, which 
agrees reasonably with the value of $|V_{ub}|$ obtained from the
inclusive end point spectrum \cite{cleo,argus}.

Alternatively, the possibility of measuring $|V_{ub}|$ via
non-leptonic decays of $B$ mesons to exclusive two meson final states
\cite{soni} has been theoretically explored.
Recently it has also been  suggested  that the measurements of
hadronic invariant mass spectrum \cite{kim-ko} as well as 
hadronic energy spectrum \cite{bouzas}
in the inclusive $B \rightarrow X_{c(u)} l \nu$ decays can be
useful in extracting $|V_{ub}|$ with better theoretical understandings.
In a future asymmetric $B$-factory with microvertex detector, 
the hadronic invariant mass spectrum will offer
alternative ways to select $b \rightarrow u$ transitions 
that are much more efficient than selecting the upper end 
region of the lepton energy spectrum, 
with much less theoretical uncertainties.
The measurement of ratio $|V_{ub}/V_{ts}|$ from the differential
decay widths of the processes $B \rightarrow \rho l \nu$ and
$B \rightarrow K^* l \bar l$ by using $SU(3)$-flavor symmetry and
the heavy quark symmetry has been also proposed \cite{sanda}.
There has also been a recent theoretical progress on the exclusive 
$b \rightarrow u$ 
semileptonic decay form factors using the HQET-based scaling laws 
to extrapolate 
the form factors from the semileptonic $D$ meson decays \cite{hqet-based}.
And their prediction is similar to most of quark model predictions 
\cite{models}.
It is urgently important that all the available methods have to be 
thoroughly explored to measure the most important CKM matrix element 
$V_{ub}$ as accurately as possible in the forthcoming $B$-factories. 
\\

\newpage
\noindent
{\bf 2. Theoretical Proposal}~~~
Over the past few years, a great progress has been achieved 
in our understanding of {\it inclusive} semileptonic decays 
of heavy mesons \cite{hqet}, especially in the lepton energy spectrum.
However, it turns out that the end-point region of the lepton energy 
spectrum cannot be described by 
$1/m_{_Q}$ expansion. Rather, a partial resummation of  $1/m_{_Q}$ 
expansion is required \cite{resum}, closely analogous to the leading
twist contribution in deep inelastic scattering, which could bring about 
significant uncertainties and presumable model dependences.

Even with a theoretical breakdown near around the end-point region of 
lepton energy spectrum, accurate prediction of the {\it total} 
integrated semileptonic decay rate can be obtained \cite{hqet} 
within the HQET including the first non-trivial, non-perturbative
corrections as well as radiative perturbative QCD correction
\cite{kuhn}. The related uncertainties in calculation of the integrated
decay rate have been also analyzed \cite{luke,shifman,ball,ural}.
The total inclusive semileptonic decay rate for 
$B \rightarrow X_q l \nu$ is given \cite{shifman} as
\begin{eqnarray}
\Gamma(B \rightarrow X_q l \nu) = 
{G_F^2 m_b^5 \over 192 \pi^3} |V_{qb}|^2
\Biggl\{ 
&\Biggl[& z_0(x_q) - {2 \alpha_s(m_b^2) \over 3 \pi}
g(x_q) \Biggr] \left( 1 - {\mu_\pi^2 - \mu_G^2 \over 2 m_b^2} \right) 
\nonumber\\
&-& z_1(x_q) {\mu_G^2 \over m_b^2} + 
{\cal O}(\alpha_s^2,\alpha_s/m_b^2,1/m_b^3)
~~\Biggr\}~~,
\label{eq2}
\end{eqnarray}
where
\begin{eqnarray}
x_q &\equiv& m_q/m_b~~, \nonumber\\
z_0(x) &=& 1 -8x^2 +8x^6 -x^8 -24x^4\log{x}~~, \nonumber\\
z_1(x) &=& (1-x^2)^4~~, \nonumber
\end{eqnarray}
and $g(x) = (\pi^2-31/4)(1-x)^2+3/2$
is the corresponding single gluon
exchange perturbative QCD correction \cite{kuhn,kim-martin}.
The expectation value of energy due to the chromomagnetic hyperfine 
interaction, $\mu_G$, can be related to the $B^* - B$ mass difference
\begin{equation}
\mu_G^2 = {3 \over 4} (M_{B^*}^2 - M_B^2) 
\approx (0.350 \pm 0.005)~{\rm GeV}^2~~,
\label{eq4}
\end{equation}
and the expectation value of kinetic energy of $b$-quark inside 
the $B$ meson, $\mu_\pi^2$, is given from various  
arguments \cite{mu-pi,kim-namgung,gremm},
\begin{equation}
0.10~{\rm GeV}^2 \leq \mu_\pi^2 \leq  0.65~{\rm GeV}^2~~,
\label{eq5}
\end{equation}
which shows much larger uncertainties compared to $\mu_G^2$.
The value of $|V_{cb}|$ has been estimated \cite{luke,shifman,ball}
from the total decay rate $\Gamma(B \rightarrow X_c l \nu)$ 
of Eq. (\ref{eq2}) by using the pole mass of 
$m_b$ and a mass difference $(m_b - m_c)$ based on the HQET. 
As can be easily seen from Eq. (\ref{eq2}), the $m_b^5$ factor,
which appears in the semileptonic decay rate, 
but not in the branching fraction, 
could be the largest source of the uncertainty, resulting in about
$5 \sim 20\%$ error in the prediction of $|V_{cb}|$ via the semileptonic 
branching fraction and $B$ meson life time.
However, we note the recent arguments \cite{shifman,ural} that 
a consistent treatment of the running masses and the perturbative QCD
correction appears to cancel the large uncertainties 
from (i) the mass term and (ii) the perturbative expansion, 
which seems to be borne out by the calculations of 
Ball $et$ $al.$ \cite{ball}.  

We can do a similar exercise to predict the value of $|V_{ub}|$ from the
integrated total decay rate of $\Gamma(B \rightarrow X_u l \nu)$, 
to find out
\begin{displaymath}
|V_{ub}|^2 = {192 \pi^3 \cdot 
\Gamma(B \rightarrow X_u l \nu) \over G_F^2 m_b^5} \Biggl\{
\left[ 1 - {2 \alpha_s(m_b^2) \over 3 \pi}
\left( \pi^2 - {25 \over 4} \right) \right]
\left( 1 - {\mu_\pi^2 - \mu_G^2 \over 2 m_b^2} \right) 
- {\mu_G^2 \over m_b^2} \Biggr\}^{-1}. 
\end{displaymath}
And by using the pole mass of $b$-quark $m_b = (4.8 \pm 0.2)$ GeV 
from a QCD sum-rule analysis\footnote{To be conservative, 
we use a larger error bar (larger by a factor 8) than that of the original 
analysis \cite{voloshin}. We estimate the largest possible error
of $m_b$ as ${\cal O}(\Lambda_{QCD})$.} 
of the $\Upsilon$-system \cite{voloshin},  $x_u \equiv m_u/m_b \simeq 0$,
and taking\footnote{Extrapolating the known 5 \% error of 
$\alpha_s(m_{_Z}^2)$, we estimate about 10 \% error for $\alpha_s(m_b^2)$.}
$\alpha_s(m_b^2) = (0.24 \pm 0.02)$, we get numerically as a conservative
estimate
\begin{eqnarray}
\gamma_{u} &\equiv&
{\Gamma_{theory}(B \rightarrow X_u l \nu) \over |V_{ub}|^2}
\simeq (7.1 \pm 1.5) \times 10^{13}/sec~~, \nonumber\\
{\rm and}~~~~~ |V_{ub}| &\simeq& (3.6 \pm 0.4) \times 10^{-3} \cdot
\left[{{\cal B}(B \rightarrow X_u l \nu) 
\over 1.4\times 10^{-3}}\right]^{1/2}
\left[ { 1.52~ psec \over \tau_{_B} } \right]^{1/2}. 
\label{eq6}
\end{eqnarray}
Note that there exists a similar estimate \cite{ural} but with smaller
error ($\sim 5\%$) by using the theoretically defined
running mass of $m_b$ normalized at the scale about 1 GeV.

We remark that the semileptonic branching fraction of 
$b \rightarrow u$ decay,
${\cal B}(B \rightarrow X_u l \nu)$, has to be precisely measured 
to experimentally determine the value of $|V_{ub}|$ from Eq. (\ref{eq6}). 
We will discuss on the
experimental possibilities in details in the next Section.
Once the inclusive branching fraction 
${\cal B}(B \rightarrow X_u l \nu)$ is 
precisely measured, we can extract the value of $|V_{ub}|$ within the 
theoretical error similar to those of $|V_{cb}|$. 

The ratio of CKM matrix elements $|V_{ub}/V_{cb}|$  can be determined in 
a model-independent way by taking the ratio of semileptonic decay widths
$\Gamma(B \rightarrow X_u l \nu)/\Gamma(B \rightarrow X_c l \nu)$.
As can be seen from Eq. (\ref{eq2}), 
this ratio is theoretically described by the 
phase space factor and the well-known perturbative QCD correction only,
\begin{equation}
{\Gamma(B \rightarrow X_u l \nu) \over \Gamma(B \rightarrow X_c l \nu)}
\simeq \left| { V_{ub} \over V_{cb} } \right|^2 
\left[ 1 - {2 \alpha_s \over 3 \pi}
\left( \pi^2 - {25 \over 4} \right) \right]
\left[ z_0(x_c) - {2 \alpha_s \over 3 \pi} g(x_c) \right]^{-1},
\label{eq7}
\end{equation}
where we ignored the term $\mu_G^2/m_b^2$, which gives about
1\% correction to the ratio.
We strongly emphasize here 
that the sources of the main possible theoretical uncertainties, 
the factor $m_b^5$ and the still-problematic non-perturbative 
contributions, are all
canceled out in this ratio. By taking $\alpha_s(m_b^2) = (0.24 \pm 0.02)$,
and by using the mass difference relation from the HQET \cite{mass}, 
which gives\footnote{This ratio $x_c$ is calculable 
from the mass difference
$(m_b-m_c)$, which also includes the uncertain parameter 
$\mu_\pi^2$ of Eq. (\ref{eq5}) as a small correction factor.}
$x_c \equiv m_c/m_b \approx 0.25 - 0.30$, 
the ratio of the semileptonic decay widths is conservatively estimated as
\begin{equation}
{\Gamma(B \rightarrow X_u l \nu) \over \Gamma(B \rightarrow X_c l \nu)}
\equiv \left( {\gamma_u \over \gamma_c} \right) \times
  \left| { V_{ub} \over V_{cb} } \right|^2
\simeq (1.83 \pm 0.28) \times \left| { V_{ub} \over V_{cb} } \right|^2,
\label{eqq8} 
\end{equation}
and the ratio of CKM elements is 
\begin{equation}
\left| { V_{ub} \over V_{cb} } \right| 
\equiv \left( {\gamma_c \over \gamma_u} \right)^{1/2} \times
\left[ {{\cal B}(B \rightarrow X_u l \nu) \over 
{\cal B}(B \rightarrow X l \nu) } \right]^{1/2}
\simeq (0.74 \pm 0.06) \times
\left[ {{\cal B}(B \rightarrow X_u l \nu) \over 
{\cal B}(B \rightarrow X_c l \nu) } \right]^{1/2}. 
\label{eq8} 
\end{equation}
We note here that within a simple spectator model of $b$-quark decay
Rosner \cite{rosner} predicted the ratio of the decay widths as
\begin{displaymath}
{\Gamma(b \rightarrow u l \nu) \over \Gamma(b \rightarrow c l \nu)} =
(1.85 \sim 2.44) \times \left|{V_{ub} \over V_{cb}}\right|^2.
\end{displaymath}
We find that this free-quark-decay estimate without including any QCD
corrections gives a rather similar result to our prediction, 
Eq. (\ref{eqq8}), based on the HQET.
Once the ratio of semileptonic decay widths (or equivalently the ratio of
branching fractions 
${\cal B}(B \rightarrow X_u l \nu)/{\cal B}(B \rightarrow X_c l \nu)$)
is measured in the forthcoming asymmetric $B$-factories, 
this should give a powerful model-independent determination 
of $|V_{ub}/V_{cb}|$.
\\

\noindent
{\bf 3. Experimental Possibility}~~~
As explained in the previous Section, in order to measure 
$|V_{ub}/V_{cb}|$ (and $|V_{ub}|$) model-independently
by using the relations, Eqs. (\ref{eq6},\ref{eqq8},\ref{eq8}), 
it is experimentally
required to separate the $b \rightarrow u$ semileptonic decays 
from the dominant $b \rightarrow c$ semileptonic decays, 
and to precisely measure the branching fraction
${\cal B}(B \rightarrow X_u l \nu)$ or the ratio
${\cal B}(B \rightarrow X_u l \nu)/{\cal B}(B \rightarrow X_c l \nu)$.

At presently existing symmetric $B$-experiments, ARGUS and CLEO, 
where $B$ and $\bar B$ are produced almost at rest, this required 
separation is possible only in the very end-point region of the lepton 
energy spectrum, because both $B$ 
and $\bar B$ decay into the whole $4 \pi$ solid angle from the almost same 
decay point, and it is not possible to identify the parent $B$ meson of 
each produced particle. Hence all the hadronic information 
is of no use. However, recently CLEO \cite{alexander} succeeded
measuring the hadronic invariant masses for the fully reconstructed
$B \rightarrow \rho l \nu$ and $B \rightarrow \omega l \nu$ decay events
by using the neutrino reconstruction technique and the beam constrained
invariant mass.
In the forthcoming asymmetric $B$-experiments with microvertex 
detectors, BABAR and BELLE \cite{B-fact}, where the two
beams have different energies and the produced $\Upsilon(4S)$ is not at
rest in the laboratory frame, the bottom decay vertices will be better
identifiable.
The efficiency for the full reconstruction of each event could be 
relatively high (maybe as large as several percentages of efficiency) 
limited  only by the $\pi^0$-reconstruction efficiency of about 
$60 \%$ \cite{B-fact}, 
and this $b \rightarrow u$ separation would be experimentally viable.

As of the most straightforward separation method, the measurements of 
inclusive hadronic invariant mass ($m_{_X}$) distributions in 
$B \rightarrow X_{c,u} l \nu$   can be very useful 
for the fully reconstructed semileptonic decay events.  
For $b \rightarrow c$ decays, one necessarily has  
$m_{_X} \geq m_{_D} = 1.86$ GeV.  
Therefore, if we impose a condition $m_{_X} < m_{_D}$, 
the resulting events come only from $b \rightarrow u$ decays,  
and about 90\% of the $b\rightarrow u$ events would survive this cut.   
This is already in sharp contrast with the usual cut 
on charged lepton energy $E_l$.
In fact, one may relax the condition\footnote{There are possibly
non-negligible contributions at $m_{_X} < m_{_{D^{**}}}$
from the broad $D^{**}$ states and/or from
$D^* \pi$ nonresonant decays.  Since we know little about the correct
hadronic mass shape in this region, we cannot subtract the 
$b \rightarrow c$ component with absolute certainty. 
The experimental smearing will further exacerbate this problem.
Therefore, it is extremely unlikely that the $m_{_X} < m_{_D}$ 
condition can be relaxed.} $m_{_X} < m_{_D}$, 
and extract almost the total $b \rightarrow u$ semileptonic decay rate 
\cite{kim-ko}, 
because the $m_{_X}$ distribution in $b \rightarrow c$ decays is completely 
dominated by contributions of three resonances $D, D^{*} $ and $D^{**}$, 
which are  essentially like $\delta$-functions, 
\begin{equation} 
{d \Gamma \over d m_{_X}} = \Gamma(B\rightarrow R l \nu)~
\delta(m_{_X} - m_{_R})~~,
\label{eq9}
\end{equation}
where the resonance $R = D, D^*$ or $D^{**}$.
In other words, one is allowed to use the $b \rightarrow u$ events 
in the region even above $m_{_X} \geq m_{_D}$, first by excluding small 
regions in $m_{_X}$ around $m_{_X} = m_{_D}, m_{_{D^{*}}}, m_{_{D^{**}}}$,
and then by including the regions again numerically 
in the $m_{_X}$ distribution of $b \rightarrow u$ decay from its values 
just around the resonances. 
There still is a non-resonant decay background 
at large invariant-mass region
$m_{_X} \geq m_{_D} + m_\pi$ from $B \rightarrow (D + \pi) l \nu$ in using 
this inclusive $m_{_X}$ distribution separation.
To avoid this non-resonant background, we have to impose a condition 
$m_{_X} < m_{_D} + m_\pi$, and we would still get
about 95\% of the total $b\rightarrow u$ semileptonic decay events.
For more details on this inclusive hadronic invariant mass distribution
$d \Gamma / d m_{_X}$, please see Ref. \cite{kim-ko}.

We would like to note the difficulties on this inclusive separation of
the $b \rightarrow u$ from the dominant $b \rightarrow c$ decays, 
when the neutral particles,
such as $K_L,~n,~\pi^0$, are produced as final decay products.
A small rate of mis-handling of these particles could lead to very long
tails on the invariant mass distribution. 
Therefore, the hadronic invariant mass has to be precisely measured, 
even for the masses well below $m_{_D}$, 
in order to separate out the true $b \rightarrow u$ decays 
from the dominant $b \rightarrow c$.  
Being able to reconstruct correctly several percentages of the events 
would be only the first step -- one must be able to suppress the
mis-reconstructed $b \rightarrow c$ events, which is 
a much harder challenge.

We also note that there is possibly a question of bias.  
Some classes of final states 
($e.g.$ those with low multiplicity, few neutrals) may be more
susceptible to a full and unambiguous reconstruction, 
as previously explained. Hence an
analysis that requires this reconstruction may be biassed. However,
the use of topological information from microvertex detectors
should tend to reduce the bias, since vertex resolvability depends
largely on the proper time of the decay and its orientation relative
to the initial momentum (that are independent of the decay mode).
Also such a bias can be allowed for in the analyses, via a suitable
Monte Carlo modeling.
There  also possibly is a source of background 
from the cascade decay of $b \rightarrow c \rightarrow s l \nu$. 
Recently ARGUS and CLEO \cite{cas-B} 
have separated this cascade decay background
from the signal events to extract the model-independent spectrum of 
${d\Gamma \over dE_l}(B \rightarrow X_c l \nu)$ for the whole region of 
electron energy, by taking care of lepton charge and $B - \bar B$ mixing 
systematically. In the future asymmetric $B$-factories with much higher 
statistics, this cascade decay may not be any serious background at all
except for the case with very low energy electron production.
We should also note 
that the decay channel 
$b \rightarrow c l \nu$ with $c \rightarrow s l \nu$
is another background source in a sense that is similar to the 
$K_L, n, \pi^0$ backgrounds mentioned earlier.  
Identifying a track as a muon 
or an electron will be problematic in the momentum 
range below 0.5 to 1 GeV.
%
This means that a large portion of the $c \rightarrow s l \nu$
decays will not have the lepton identified.  
These events can thus appear 
experimentally as single-lepton events with a low hadronic mass.   
This kind of semileptonic cascade background where the secondary lepton 
is not identified would also be very serious problem unless the
experimentalists find the solution to avoid systematically.  
\\

\noindent
{\bf 4. Summary}~~~
The precise value of $V_{ub}$ is urgently needed 
for understanding the origin of CP-violation, for
testing the SM through the unitarity relations
among them, and also in the searches for hints of new physics 
beyond the SM.
We propose that the ratio of CKM matrix elements $|V_{ub}/V_{cb}|$  
can be determined in a model-independent way by taking the ratio 
of semileptonic decay widths
$\Gamma(B \rightarrow X_u l \nu)/\Gamma(B \rightarrow X_c l \nu)$, 
which is theoretically described by the phase space factor
and the well-known perturbative QCD correction only, 
and which is conservatively estimated as
\begin{displaymath}
{\Gamma(B \rightarrow X_u l \nu) \over \Gamma(B \rightarrow X_c l \nu)}
\equiv \left( {\gamma_u \over \gamma_c} \right) \times
  \left| { V_{ub} \over V_{cb} } \right|^2
\simeq (1.83 \pm 0.28) \times \left| { V_{ub} \over V_{cb} } \right|^2,
\end{displaymath}
and  
\begin{displaymath}
\left| { V_{ub} \over V_{cb} } \right| 
\equiv \left( {\gamma_c \over \gamma_u} \right)^{1/2} \times
\left[ {{\cal B}(B \rightarrow X_u l \nu) \over 
{\cal B}(B \rightarrow X l \nu) } \right]^{1/2}
\simeq (0.74 \pm 0.06) \times
\left[ {{\cal B}(B \rightarrow X_u l \nu) \over 
{\cal B}(B \rightarrow X_c l \nu) } \right]^{1/2}, 
\end{displaymath}
based on the heavy quark effective theory.
Once the ratio of semileptonic decay widths 
(or equivalently the ratio of branching fractions 
${\cal B}(B \rightarrow X_u l \nu)/{\cal B}(B \rightarrow X_c l \nu)$)
is measured, this ratio will give a powerful 
model-independent determination of $|V_{ub}/V_{cb}|$.

In the forthcoming asymmetric $B$-factories with microvertex 
detectors, the total separation of $b \rightarrow u$ 
semileptonic decays from the dominant $b \rightarrow c$ semileptonic 
decays to determine the ratio would be experimentally viable.
We explore the possible experimental option: the measurement of
inclusive hadronic invariant mass distributions.
We also clarify the relevant experimental backgrounds.
In view of the potential importance of 
${\cal B}(B \rightarrow X_u l \nu )/{\cal B}(B \rightarrow X_c l \nu )$ 
as a new theoretically model-independent probe for measuring
$|V_{ub}/V_{cb}|$,
we would like to urge our experimental colleagues to make sure that 
this $b \rightarrow u$ separation can indeed be successfully achieved.

\begin{center}
{\Large Acknowledgements}
\end{center}

\noindent
The work  was supported 
in part by the Basic Science Research Institute Program,
Ministry of Education 1997,  Project No. BSRI-97-2425,
in part by CTP of SNU, 
in part by Yonsei University Faculty Research Fund of 1997,
and in part by KOSEF-DFG large collaboration
project, Project No. 96-0702-01-01-2.

\pagebreak

\end{document}